# EFFICIENT MODELLING & FORECASTING WITH RANGE BASED VOLATILITY MODELS AND APPLICATION

Ng, Kok Haur[1]
[1]Institute of Mathematical Sciences, Faculty of Science, University of Malaya, Malaysia.

Shelton Peiris[2]; Jennifer, So-kuen Chan[3]; David Allen[5]
[2,3,5]School of Mathematics and Statistics, Faculty of Science, The University of Sydney, Australia

Ng, Kooi-Huat [4]
[4]Department of Mathematical and Actuarial Sciences, Universiti Tunku Abdul Rahman, Malaysia.

**Abstract**

*This paper considers an alternative method for fitting CARR models using combined estimating functions (CEF) by showing its usefulness in applications in economics and quantitative finance. The associated information matrix for corresponding new estimates is derived to calculate the standard errors. A simulation study is carried out to demonstrate its superiority relative to other two competitors: linear estimating functions (LEF) and the maximum likelihood (ML). Results show that CEF estimates are more efficient than LEF and ML estimates when the error distribution is mis-specified. Taking a real data set from financial economics, we illustrate the usefulness and applicability of the CEF method in practice and report reliable forecast values to minimize the risk in the decision making process.*

# 1. Introduction

Autoregressive Conditional Heteroskedastic (ARCH), Generalized ARCH (GARCH) and stochastic volatility models are widely used in modelling the dynamics of volatility in financial markets. However, many studies including Alizadeh et al. (2002), Chou (2005), Brandt and Diebold (2006) have suggested that GARCH models are not always accurate and efficient in certain applications. A reason for this is that they are return-based and use only the data available at closing prices ignoring the intra-day price movements and related information. As a result, the range, defined as being the difference between high and low prices over a given time interval, becomes a popular alternative measure, since it utilizes two pieces of information (the high and low prices) from the market within the given interval. Alizaded et al. (2002) showed that range-based models are more efficient than traditional volatility models in volatility or risk modelling.

As an alternative approach, Chou (2005) proposed a symmetric range-based model called Conditional Autoregressive Range (CARR) model and an extension of the CARR model with exogenous variables called the CARRX model to forecast volatilities. He claimed that the CARR model provides a simple, efficient and natural framework to analyze the dynamics of volatility. Empirical results also show that the CARR model gives sharper volatility estimates than a standard GARCH model. Brandt and Diebold (2006), Brandt and Jones (2006) and Lin et al. (2012) also considered range-based models on various financial markets and their applications can be found in Chou and Liu (2010), Li and Hong (2011), Chou et al. (2013) and Sin (2013). Further extensions of CARR models include the threshold conditional autoregressive model of Chen et al. (2008), the time-varying logarithmic conditional autoregressive range model of Chiang and Wang (2011), the conditional autoregressive geometric process range model of Chan et al. (2012) and the CARR model incorporating the sudden change components of Kumar (2015).

CARR models possess similar features to autoregressive conditional duration (ACD) models except that the range data measures the maximum change of price over fixed time intervals while the duration data measures the time interval between two consecutive transactions. As pointed out by Allen et al. (2013), two important issues in the use of ACD models are model specification and the choice of a suitable distribution for the errors. Since the exact distribution of errors is often unknown in practice, previous researches on CARR and ACD models consider mainly parametric models with extensions to more flexible error distributions such as log-t (Chan et al., 2012). Recently, Andres and Harvey (2012) considered the generalized beta type 2 (GB2) distribution in financial modelling as it nests many important distributions as special cases. However, the estimation of parameters for GB2 distribution can be difficult due its complicated density function. To avoid such difficulties in the maximum likelihood (ML) approach, alternative approaches have been adopted for CARR models. For example, Chou (2005) and Chou et al. (2013) considered the quasi ML method, Chiang and Wang (2011) applied a two-stage ML method while Chen et al. (2008) and Chan et al. (2012) used the Bayesian approach.

However, it is not surprising that a flexible distribution such as GB2 may still fail to model some data well. A semi-parametric approach using the theory of estimating functions (EF) avoids any distributional assumption and hence can be applied to more general data sets. The linear estimating function (LEF) method has been successfully applied in many cases of financial time series modeling. See, for example, Bera et al. (2006) and Allen et al. (2013). The LEF method was further extended to include combined estimating functions (CEF) (Li and Turtle, 2000) to incorporate with higher order moments and applied to ARCH and GARCH models. Liang et al. (2011) introduced a general framework for joint estimation of time series models using the quadratic estimating functions (QEF) which shares the same approach to CEF and applied it to an ACD model with a doubly stochastic model, a random coefficient autoregressive model and a regression model with ARCH errors. Meanwhile, Ng et al. (2015) developed the QEF procedure for log-ACD models.

This paper considers the CEF method to estimate parameters of CARR models. Estimating functions and closed form expressions for information matrices related to CEF and LEF estimators are derived and the corresponding statistical properties are validated through theoretical and simulation results. Parameter estimations of CARR models with known as well as mis-specified error distributions based on CEF, LEF, ML methods are compared and reported in this study. These estimators are examined and assessed through the GB2 distribution, including error mis-specifications. Finally, the applicability of the CEF method is also demonstrated via an empirical example.

The remainder of this paper is structured as follows: Section 2 reviews the use of the ML, LEF and CEF methods. Section 3 investigates the performance of these three estimators through a simulation study. Section 4 reports an empirical application to illustrate the use of the CEF method for the CARR models and to provide an assessment of the forecasting abilities, and finally concluding remarks are given in Section 5.

## 2. Estimating CARR Models

Suppose that $P_{t\tau}$ is the logarithmic price of a speculative asset at time $\tau$ of the $t$-th fixed interval, where $\tau = 1,2,\ldots,n_t$; $t = 1,2,\ldots,T$. The range of $P_{t\tau}$ over a fixed interval $t$ (e.g. a day) measures the dispersion across $\tau$. The range of $P_{t\tau}$ for a fixed $t$ is defined as

$$r_t = 100 \times \left( \max_{1 \leq \tau \leq n_t} \{P_{t\tau}\} - \min_{1 \leq \tau \leq n_t} \{P_{t\tau}\} \right); \ t = 1,2,\ldots,T. \tag{1}$$

The CARR $(u,v)$ model for the range is defined as

$$r_t = \psi_t \varepsilon_t,$$

where $\{\varepsilon_t\}$ is an independent and identically distributed sequence such that $E(\varepsilon_t) = \mu_\varepsilon$ and $\text{Var}(\varepsilon_t) = \sigma_\varepsilon^2$. Its skewness and excess kurtosis are denoted by $\gamma_\varepsilon$ and $\kappa_\varepsilon$ respectively. Further, the dynamic specification of $\psi_t$ is given by

$$\psi_t = \omega + \sum_{i=1}^{u} \alpha_i r_{t-i} + \sum_{j=1}^{v} \beta_j \psi_{t-j},$$

where $\omega$, $\alpha_i$ and $\beta_j$ ($\omega > 0$, $\alpha_i, \beta_j \geq 0$) are parameters to be estimated.

### 2.1 Method of Maximum Likelihood (MLE)

In practice, it is advisable to use a very flexible distribution for the errors $\varepsilon_t$. One choice is the GB2 distribution which nests many important distributions such as the gamma, Weibull, Pareto, Burr12, lognormal, generalized gamma and the Pearson family, as special cases. The probability density function (pdf) of GB2 for $\varepsilon_t$ is given by

$$f(\varepsilon_t \mid a, b_t, p, q) = \frac{|a| \varepsilon_t^{ap-1}}{b_t^{ap} B(p,q)[1 + (\varepsilon_t / b_t)^a]^{p+q}}, \ \varepsilon_t > 0, \tag{2}$$

where $a, p$ and $q$ are the shape parameters such that $p, q > 0$ and $a \neq 0$, $b_t (> 0)$ is a scale parameter depending on $t$, and $B(p,q)$ is the beta function. It is easy to verify that the mean of the distribution is given by

$$E(\varepsilon_t) = \frac{b_t B(p + \frac{1}{a}, q - \frac{1}{a})}{B(p,q)}. \tag{3}$$

See Cummins et al. (1990), pp. 260 for details.

To derive the ML estimator, we consider the log-likelihood function $l(\boldsymbol{\lambda})$:

$$l(\boldsymbol{\lambda}) = \sum_{t=1}^{T} \log f(r_t | \boldsymbol{\lambda}), \tag{4}$$

where $\boldsymbol{\lambda}$ is the vector of all parameters $(a, p, q, \omega, \alpha_1, \alpha_2, ..., \alpha_u, \beta_1, \beta_2, ..., \beta_v)$ and the density $f(r_t | \boldsymbol{\lambda})$ can be obtained using Eq. (2).

The corresponding ML estimator, $\hat{\boldsymbol{\lambda}}$, of $\boldsymbol{\lambda}$ is evaluated by

$$\hat{\boldsymbol{\lambda}} = \arg \max_{\boldsymbol{\lambda} \in \Lambda} l(\boldsymbol{\lambda}), \tag{5}$$

where $\Lambda$ represents the vector space of all parameters and $\hat{\boldsymbol{\lambda}}$ can be obtained using a suitable optimization algorithm, for example, Optim in the R library package or fminsearch in Matlab.

### 2.2 Linear Estimating Functions (LEF)

As the true distribution of $\varepsilon_t$ is seldom known in practice, a semi-parametric method avoids the exact specification of the error distribution.

Suppose that $\{r_1, r_2, ..., r_T\}$ is a discrete valued stochastic process and we are interested in fitting a suitable model for this sample. Let $\Theta$ be the class of probability distributions $F$ on $\Re^T$ and $\boldsymbol{\theta} = \boldsymbol{\theta}(F)$ be a vector of real parameters for $F \in \Theta$. Define $E_F(\cdot)$ and $E_{t-1,F}(\cdot)$ to be the conditional expectations with respect to $F$ and the later by further holding the first $t-1$ values $r_1, r_2, ..., r_{t-1}$ fixed. For simplicity, we write $E_{t-1,F}(\cdot) = E_{t-1}(\cdot)$ and $E_F(\cdot) = E(\cdot)$. Further, let $h_t(\cdot) = h_t$ be a real valued function of $r_1, r_2, ..., r_t$ and parameters $\boldsymbol{\theta}$ such that

$$E_{t-1}[h_t] = 0, \quad (t = 1, 2, ..., T; F \in \Theta), \tag{6}$$

and

$$E_{t-1}[h_t h_s] = 0, \, t \neq s. \tag{7}$$

Assuming that $\mathbf{g}(\mathbf{r}; \boldsymbol{\theta})$ are real valued functions of the random variables $\mathbf{r} = \{r_1, r_2, ..., r_T\}$ and the parameters $\boldsymbol{\theta}$ fulfill the standard regularity conditions (see e.g. Godambe, 1985). Then the function $\mathbf{g}(\mathbf{r}; \boldsymbol{\theta})$ satisfying $E[\mathbf{g}(\mathbf{r}; \boldsymbol{\theta})] = \mathbf{0}$ is called a regular unbiased estimating function. Among all regular unbiased estimating functions $\mathbf{g}(\mathbf{r}; \boldsymbol{\theta})$, $\mathbf{g}^*(\mathbf{r}; \boldsymbol{\theta})$ is said to be optimum if

$$E[\mathbf{g}(\mathbf{r};\boldsymbol{\theta})\mathbf{g}(\mathbf{r};\boldsymbol{\theta})'] \left\{ E\left(\frac{\partial \mathbf{g}(\mathbf{r};\boldsymbol{\theta})}{\partial \boldsymbol{\theta}}\right) E\left(\frac{\partial \mathbf{g}(\mathbf{r};\boldsymbol{\theta})}{\partial \boldsymbol{\theta}}\right)' \right\}^{-1}, \tag{8}$$

is minimized for all $F \in \Theta$ at $\mathbf{g}(\mathbf{r};\boldsymbol{\theta}) = \mathbf{g}^*(\mathbf{r};\boldsymbol{\theta})$. An optimal estimate of $\boldsymbol{\theta}$ is obtained by solving the optimum estimating equation $\mathbf{g}^*(\mathbf{r};\boldsymbol{\theta}) = \mathbf{0}$.

Consider the class of linear unbiased estimating functions $L$ formed by

$$\mathbf{g}(\mathbf{r};\boldsymbol{\theta}) = \sum_{t=1}^{T} a_{t-1} h_t, \tag{9}$$

where $a_{t-1}$ is a suitably chosen function of the random variables $r_1, r_2, \cdots, r_{t-1}$ and the parameters $\boldsymbol{\theta}$ for all $t = 1, 2, \ldots, T$. From Eqs. (6) and (9), it is apparent that, $E[\mathbf{g}(\mathbf{r};\boldsymbol{\theta})] = \mathbf{0}$ for all $\mathbf{g}(\mathbf{r};\boldsymbol{\theta}) \in L$. Following the optimal theorem of Godambe (1985), the function $\mathbf{g}^*(\mathbf{r};\boldsymbol{\theta})$ minimizing Eq. (8) is given by

$$\mathbf{g}^*(\mathbf{r};\boldsymbol{\theta}) = \sum_{t=1}^{T} a_{t-1}^* h_t, \tag{10}$$

where $a_{t-1}^* = \left( E_{t-1}\left[\frac{\partial h_t}{\partial \boldsymbol{\theta}}\right] \right) \left( E_{t-1}[h_t^2] \right)^{-1}$.

An optimal estimate of $\boldsymbol{\theta}$ (in the sense of Godambe, 1985) is acquired by solving the optimal equation(s) $\mathbf{g}^*(\mathbf{r};\boldsymbol{\theta}) = \mathbf{0}$.

Supposing that the errors $\varepsilon_t$ for a CARR model follow a distribution with a positive support, it can be shown that the mean and variance of the conditional distribution of $r_t$ given $F_{t-1}$ are $E(r_t | F_{t-1}) = \psi_t \mu_\varepsilon$ and $\text{Var}(r_t | F_{t-1}) = \psi_t^2 \sigma_\varepsilon^2$ respectively. To find the LEF estimates for the CARR model, we define $h_t = r_t - \psi_t \mu_\varepsilon$ to be a martingale difference sequence. It can be easily verified that $h_t$ fulfils two regularity conditions, namely, the unbiasedness in Eq. (6) and the mutual orthogonality in Eq. (7). Hence the optimal LEF of Eq. (10) becomes

$$\mathbf{g}_1^*(\mathbf{r};\boldsymbol{\theta}) = -\sum_{t=1}^{T} \frac{\mu_\varepsilon}{\psi_t^2 \sigma_\varepsilon^2} \frac{\partial \psi_t}{\partial \boldsymbol{\theta}} (r_t - \psi_t \mu_\varepsilon), \tag{11}$$

since

$$a_{t-1}^* = \left( E_{t-1}\left[\frac{\partial h_t}{\partial \boldsymbol{\theta}}\right] \right) \left( E_{t-1}[h_t^2] \right)^{-1} = -\mu_\varepsilon \frac{\partial \psi_t}{\partial \boldsymbol{\theta}} \bigg/ \psi_t^2 \sigma_\varepsilon^2,$$

and $\boldsymbol{\theta} = (\omega, \alpha_1, \alpha_2, \ldots, \alpha_u, \beta_1, \beta_2, \ldots, \beta_v)$. An optimal estimate of $\boldsymbol{\theta}$ can be obtained by solving the equation(s) $\mathbf{g}_1^*(\mathbf{r}; \boldsymbol{\theta}) = \mathbf{0}$ and the corresponding information matrix $\mathbf{I}_{\mathbf{g}_1^*}$ is given by

$$\mathbf{I}_{\mathbf{g}_1^*} = \sum_{t=1}^{T} \frac{\mu_\varepsilon^2}{\psi_t^2} \frac{1}{\sigma_\varepsilon^2} \left( \frac{\partial \psi_t}{\partial \boldsymbol{\theta}} \right) \left( \frac{\partial \psi_t}{\partial \boldsymbol{\theta}'} \right). \tag{12}$$

### 2.3 Combined Estimating Functions (CEF)

While the LEF considers only one single martingale difference sequence $h_t = r_t - \psi_t \mu_\varepsilon$, the CEF method extends the LEF method by utilizing two martingale differences $h_{t,1} = r_t - \psi_t \mu_\varepsilon$ and $\xi_t = (r_t - \psi_t \mu_\varepsilon)^2 - \psi_t^2 \sigma_\varepsilon^2$ and hence is shown to be an improvement on LEF. Since $h_{t,1}$ and $\xi_t$ are not mutually orthogonal, we use a similar approach given in Li and Turtle (2000) to define the function

$$h_{t,2} = (r_t - \psi_t \mu_\varepsilon)^2 - \psi_t^2 \sigma_\varepsilon^2 - m_t (r_t - \psi_t \mu_\varepsilon), \tag{13}$$

where $m_t$ is an arbitrary function of $t$. We aim to find $m_t$ to achieve mutual orthogonality of $h_{t,1}$ and $h_{t,2}$ such that $E_{t-1}(h_{t,1} h_{t,2}) = 0$. Hence, the required function in Eq. (13) becomes

$$h_{t,2} = (r_t - \psi_t \mu_\varepsilon)^2 - \psi_t^2 \sigma_\varepsilon^2 - \gamma_t (\psi_t \sigma_\varepsilon)(r_t - \psi_t \mu_\varepsilon) \tag{14}$$

where $\gamma_t$ is the conditional skewness of $r_t$. Then the following CEF is formulated:

$$\mathbf{g}_2^*(\mathbf{r}; \boldsymbol{\theta}) = \sum_{t=1}^{T} a_{t-1,1}^* h_{t,1} + \sum_{t=1}^{T} a_{t-1,2}^* h_{t,2}, \tag{15}$$

to estimate parameters of a CARR model where

$$a_{t-1,1}^* = \left( E_{t-1}\left[ \frac{\partial h_{t,1}}{\partial \boldsymbol{\theta}} \right] \right) \left( E_{t-1}[h_{t,1}^2] \right)^{-1} = -\mu_\varepsilon \frac{\partial \psi_t}{\partial \boldsymbol{\theta}} \bigg/ \psi_t^2 \sigma_\varepsilon^2,$$

$$a_{t-1,2}^* = \left( E_{t-1}\left[ \frac{\partial h_{t,2}}{\partial \boldsymbol{\theta}} \right] \right) \left( E_{t-1}[h_{t,2}^2] \right)^{-1} = \left( \gamma_t \sigma_\varepsilon \mu_\varepsilon \psi_t \frac{\partial \psi_t}{\partial \boldsymbol{\theta}} - 2\sigma_\varepsilon^2 \psi_t \frac{\partial \psi_t}{\partial \boldsymbol{\theta}} \right) \bigg/ \psi_t^4 \sigma_\varepsilon^4 (\kappa_t + 2 - \gamma_t^2),$$

and the corresponding excess kurtosis is defined by $\kappa_t = \dfrac{E[(r_t - \psi_t \mu_\varepsilon)^4 \mid F_{t-1}]}{(\psi_t \sigma_\varepsilon)^4} - 3$. An optimal estimate of $\boldsymbol{\theta}$ can be obtained by solving the equation(s) $\mathbf{g}_2^*(\mathbf{r}; \boldsymbol{\theta}) = \mathbf{0}$.

It is not difficult to show that the conditional skewness and excess kurtosis of $r_t$ are equal to the skewness and excess kurtosis of $\varepsilon_t$ respectively, that is,

$$\gamma_t = \gamma_\varepsilon \text{ and } \kappa_t = \kappa_\varepsilon. \tag{16}$$

The information matrix for the corresponding CEF estimates is

$$\mathbf{I}_{\mathbf{g}_2^*} = \sum_{t=1}^{T} \frac{\mu_\varepsilon^2}{\sigma_\varepsilon^2} \frac{1}{\psi_t^2} \left(\frac{\partial \psi_t}{\partial \boldsymbol{\theta}}\right) \left(\frac{\partial \psi_t}{\partial \boldsymbol{\theta}'}\right) + \sum_{t=1}^{T} \frac{(\gamma_\varepsilon \mu_\varepsilon - 2\sigma_\varepsilon)^2}{\psi_t^2 \sigma_\varepsilon^2 (\kappa_\varepsilon + 2 - \gamma_\varepsilon^2)} \left(\frac{\partial \psi_t}{\partial \boldsymbol{\theta}}\right) \left(\frac{\partial \psi_t}{\partial \boldsymbol{\theta}'}\right). \tag{17}$$

It is clear that $\mathbf{I}_{\mathbf{g}_2^*}$ is more informative than $\mathbf{I}_{\mathbf{g}_1^*}$ since

$$\mathbf{I}_{\mathbf{g}_2^*} - \mathbf{I}_{\mathbf{g}_1^*} = \sum_{t=1}^{T} \frac{(\gamma_t \mu_\varepsilon - 2\sigma_\varepsilon)^2}{\psi_t^2 \sigma_\varepsilon^2 (\kappa_\varepsilon + 2 - \gamma_\varepsilon^2)} \left(\frac{\partial \psi_t}{\partial \boldsymbol{\theta}}\right) \left(\frac{\partial \psi_t}{\partial \boldsymbol{\theta}'}\right) \geq 0. \tag{18}$$

These results can also be obtained through the QEF method. In the next section, a large scale simulation study is conducted to verify the performance of the CEF estimator relative to its competitors, the ML and LEF estimators.

### 3. Simulation Study

A simulation study is carried out using the CARR (1,1) model given by

$$r_t = \psi_t \varepsilon_t,$$

where

$$\psi_t = \omega + \alpha_1 r_{t-1} + \beta_1 \psi_{t-1}, \tag{19}$$

and the errors $\varepsilon_t$ follow a standardized GB2 distribution. The GB2 distribution is chosen to simulate the errors because it can provide more general distribution for wider applications. From Eq. (3), the scale parameter $b_t$ is set to be

$$b_t = \frac{B(p,q)}{B(p + \frac{1}{a}, q - \frac{1}{a})},$$

so that $E(\varepsilon_t) = 1$ and the pdf for the range $r_t$ is given by

$$f(r_t \mid a, p, q, \psi_t) = \frac{|a| \left(\frac{r_t}{\psi_t}\right)^{ap-1} B(p + 1/a, q - 1/a)^{ap}}{\psi_t B(p,q)^{ap+1} [1 + (\frac{r_t B(p+1/a, q-1/a)}{\psi_t B(p,q)})^a]^{p+q}}. \tag{20}$$

We first simulate a time series data of length $T$ using the above CARR model with $\boldsymbol{\theta} = (\omega, \alpha_1, \beta_1) = (0.2, 0.3, 0.4)$, $\psi_1 = 0.5$, and the errors $\varepsilon_t$ follows a standardized GB2 distribution with $a = 1.0$, $p = 1.0$, $q = 2.0$. Using this sample,

- LEF estimates are obtained by solving $\mathbf{g}_1^*(\mathbf{r}; \boldsymbol{\theta}) = \mathbf{0}$ in Eq. (11) for $\sigma_\varepsilon$ and $(\omega, \alpha_1, \beta_1)$.

- CEF estimates are obtained by solving $\mathbf{g}_2^*(\mathbf{r};\boldsymbol{\theta})=\mathbf{0}$ in Eq. (15) for $\sigma_\varepsilon$ and $(\omega,\alpha_1,\beta_1)$ with $\mu_\varepsilon=1$, $\hat{\gamma}_\varepsilon=\gamma_r$ and $\hat{\kappa}_\varepsilon=\kappa_r$, where $\gamma_r$ and $\kappa_r$ denote the sample skewness and excess kurtosis respectively (Li and Turtle, 2000).
- ML estimates are obtained by maximizing the log-likelihood function based on Eq. (20).

This procedure is repeated $N=2000$ times and for $T=500, 1000, 1500, 2000$. Then for each $T$, the mean, bias, standard deviation (SD) and root mean squared error (RMSE) being the average over the replicated data sets for each parameter are computed. To further assess the variability of estimators, the information matrices given in Eqs. (12) and (17) are computed and the standard errors (SE) (being the square root of the diagonal entries of their inverses averaged over replications) are reported. Table 1 reports the corresponding results.

From Table 1, we observe the average estimates produced by all three methods are fairly close to the true values. Obviously, the performance of all parameter estimates improves with increasing $T$ according to bias, SD and RMSE. Not surprisingly, the ML method assuming the true GB2 error distribution provides the best estimates according to all criteria in general. However, we also notice that CEF estimates are only marginally inferior to ML estimates when $T=2000$ but are comparable to ML estimates when $T=500$. Moreover the CEF estimates are substantially better than LEF estimates.

Figures 1-3 present the histograms for $\hat{\omega},\hat{\alpha}_1$ and $\hat{\beta}_1$ respectively in Table 1 using the CEF method when $T=2000$. The theoretical normal curves, with the true values as centers and SE as standard deviations are added to facilitate the comparison. The results show that the finite sample marginal distributions of $\hat{\omega},\hat{\alpha}_1$ and $\hat{\beta}_1$ closely approximate the theoretical Gaussian distributions with means 0.2, 0.3 and 0.4 as their true values. This finding is further confirmed by the close agreement between SD and SE in Table 1 using the CEF as well as the LEF methods.

Although the ML method performs well when the distributions are known, semi-parametric methods such as LEF and CEF are robust and preferable since the distributions are unknown in practice. To investigate this, a simulation study was carried-out with the three proposed methods and error mis-specification. Again, we simulate a time series of length $T$ using the parameters $\boldsymbol{\theta}=(\omega,\alpha_1,\beta_1)=(0.2,0.3,04)$, $\psi_1=0.5$ and the errors $\varepsilon_t$ from the standard lognormal distribution with $\sigma=0.5$. We note that the ML estimates are obtained by maximizing the log-likelihood function based on GB2 distribution with the pdf in Eq. (20). The procedure is repeated $N = 2000$ times for each $T = 500$, 1000, 1500, 2000 and the corresponding statistics for each estimates are given in Table 2. From the Table 2, we observe that the CEF estimates give smaller bias, SD and RMSE than the LEF and ML estimates for all sample sizes except $T = 1000$.

In comparisons of the model robustness and efficiency, we conclude that the CEF method outperforms the other two methods and hence is preferred in practice.

### 4. An Application from Financial Economics

Consider the price data from the All Ordinaries (AORD) index from the Australian market during 1 May 2009 to 1 May 2015 was used for the empirical analysis to illustrate the applicability of CARR models using the CEF estimator. The daily range $r_t$ with 1514 observations were calculated using Eq. (1). Table 3 reports their summary statistics and the Ljung-Box statistics $Q_6$ and $Q_{12}$ which test the overall randomness of $r_t$ based on 6 and 12 lagged autocorrelations respectively. As they are significant at a 5% significance level, the series of $r_t$ is non-random and hence the CARR model is adopted to account for the autocorrelation. Figures 4 and 5 display the time series plot of $r_t$ and the histogram of the series, respectively. The autocorrelation plot in Figure 6 also confirms the high level of autocorrelation for $r_t$.

### 4.1 Numerical Results

We fit a CARR (1,1) model to the data using ML, LEF and CEF methods given in Sections 2.1, 2.2 and 2.3, respectively, for the first $T = 1500$ data. All parameter estimates and their standard errors for all models are presented in Table 4. It can be seen that parameter estimates based on the three methods are very similar. Moreover, the predictive performance according to in-sample root mean square prediction error (RMSPE) and mean absolute prediction error (MAPE) are comparable but the standard errors for parameter estimates using the CEF method are much lower than those using LEF and ML methods. Therefore, the CEF method is preferred to the LEF and ML methods.

Subsequently, higher order CARR models including CARR (1,2), CARR (2,1) and CARR (2,2) models are also fitted via the CEF method. The results of the fitted models are given in Table 5. Although CARR (2,1) model yields comparable in-sample fit and better out-of-sample forecasts among all models, its parameter estimate $\hat{\alpha}_2$ which signifies CARR(2,1) model is not significant and hence it shares essentially the same structure as CARR(1,1) model. In summary, CARR (1,1) model provides the best performance among all models for both in-sample prediction and out-of-sample forecast and hence is chosen to analyse this data.

Figure 7 shows the fitted time series superimposed on the observed series. The results show that the CARR (1,1) model using the CEF method captures the general trend, persistence and volatility clustering well. We estimate the unconditional mean and variance of CARR (1,1) using

$$\mu_r = \omega/(1-\alpha_1-\beta_1),$$

and

$$\sigma_r^2 = [E(\varepsilon_t^2)-1]\mu_r^2(1-\beta_1^2-2\alpha_1\beta_1)/[1-\alpha_1^2 E(\varepsilon_t^2)-\beta_1^2-2\alpha_1\beta_1],$$

respectively, where $E(\varepsilon_t^2) = \hat{\sigma}_\varepsilon^2 + 1 = 1.1876$. We have the following results:

- ML method: $\hat{\mu}_r = 0.9781$ and $\hat{\sigma}_r^2 = 0.2556$
- LEF method: $\hat{\mu}_r = 0.9772$ and $\hat{\sigma}_r^2 = 0.2513$
- CEF method: $\hat{\mu}_r = 0.9779$ and $\hat{\sigma}_r^2 = 0.2532$

These numbers are close to the sample mean of 0.9826 and the sample variance of 0.2568. Therefore all fitted models seem adequate.

### 4.2 Forecasting

Based on the chosen CARR (1,1) models and the last $m = 14$ range data, we obtain 14-step ahead forecasts $\hat{r}_t = E(r_t) = \hat{\psi}_t$ using each of the three methods. We compare the forecast performance of these models using two out-of-sample criteria, namely the out-of-sample root mean square forecast error (RMSFE) and out-of-sample mean absolute forecast error (MAFE). The results in Table 4 indicate that the three methods provide similar forecasting performance despite the fact that the CEF method marginally outperforms the LEF and ML methods.

Although the forecast performance is similar across the three methods, the forecasts using CEF method is preferred as CEF method provides more precise parameter estimates, an advantage over the other methods. We calculate the conditional forecast variance based on:

$$Var(\hat{r}_t) = \hat{\mu}_{\psi_t}^2 \hat{\sigma}_\varepsilon^2 + \hat{\sigma}_{\psi_t}^2 + \hat{\sigma}_{\psi_t}^2 \hat{\sigma}_\varepsilon^2,$$

where $\hat{\sigma}_\varepsilon = 0.4286$ is given in Table 4 and $\hat{\sigma}_{\psi_t}^2 = 0.0056$ is calculated using the estimated information matrix from Eq. (17) where $\hat{\psi}_t$ is given in Eq. (19). Then the 95% confidence limits for the forecast $\hat{r}_t$ are

$$\hat{\psi}_t \pm 1.96\sqrt{Var(\hat{r}_t)}.$$

The forecasts using CARR (1,1) model and CEF method are plotted in Figure 8 together with their 95% confidence intervals. Figure 8 shows that the forecasts follow the trend of the observed range well. We also obtain that the coverage probability is 0.9286 which is very closed to the theoretical value of 0.95.

## 5. Concluding Remarks

The price range was shown to be an efficient risk measure in financial markets. This paper proposed the CEF method to estimate the CARR model for price range data. The superiority of the CEF estimator relative to the LEF and ML estimators are verified. The results show that the CEF estimates provide smaller bias, SD and RMSE than the LEF and ML estimates, especially when the error distribution is mis-specified which is actually common in practice. On the other hand, estimators obtained using the CEF and LEF methods are free from any distributional assumption.

The applicability of the CEF method is demonstrated using an empirical example based on the daily range data from the AORD. Its predictive power is demonstrated using the in-sample and out-of-sample predictive performance criteria and the fitted line plot. Even though these results show that the predictive and forecast performances based on CEF method is marginally outperforms the LEF and ML methods but the standard errors of parameter estimates are much lower compared to those using LEF and ML methods.

**Acknowledgement**

This work is partially supported by the UMRG grant RG260-13AFR from the University of Malaya. The first author acknowledge the support from the School of Mathematics and Statistics at the University of Sydney during his visit in 2016.

**References**

[1] Andres, P., Harvey, A. (2012). The Dynamic Location/scale Model: With Applications to Intra-day Financial Data. Cambridge Working Papers in Economics, CWPE1240, University of Cambridge.

[2] Alizadeh, S., Brandt, M.W., Diebold, F.X. (2002). Range-based of Stochastic Volatility Models or Exchange Rate Dynamics are More Interesting than You Think. *Journal of Finance*, 57(3): 1047-1092.

[3] Allen, D., Ng, K.H., Peiris, S. (2013). Estimating and Simulating Weibull Models of Risk or Price Durations: An Application to ACD Models. *North American Journal of Economics and Finance*, 25: 214-225.

[4] Bera, A. K., Bilias, Y., Simlai, P. (2006). Estimating Functions and Equations: An Essay on Historical Developments with Applications to Economics. In: Mills, T.C., Patterson, K. (Eds.), Palgrave Handbook of Econometrics, Vol. 1. pp 427 -476.

[5] Brandt, M. W., Diebold, F.X. (2006). A No-arbitrage Approach to Range-based Estimation of Return Covariances and Correlations. *Journal of Business*, 79(1): 61-73.

[6] Brandt, M. W., Jones, C. S. (2006). Volatility Forecasting with Range-based EGARCH Models. *Journal of Business & Economic Statistics*, 24(4): 470-486.

[7] Chan, J.S.K., Lam, C.P.Y., Yu, P.L.H., Choy, S.T.B., Chen, C.W.S. (2012). A Bayesian Conditional Autoregressive Geometric Process Model for Range Data. *Computational Statistics and Data Analysis*, 56(11): 3006-3019.

[8] Chen, C.W.S., Gerlach, R., Lin, E.M.H. (2008). Volatility Forecasting using Threshold Heteroskedastic Models of the Intra-day Range. *Computational Statistics and Data Analysis*, 52(6): 2990-3010.

[9] Chiang, M.H., Wang, L.M. (2011). Volatility Contagion: A Range-based Volatility Approach. *Journal of Econometrics*, 165(2): 175-189.

[10] Chou, R.Y. (2005). Forecasting Financial Volatilities with Extreme Values: The Conditional Autoregressive Range (CARR) Model. *Journal of Money Credit and Banking*, 37(3): 561-582.

[11] Chou, R.Y., Liu, N. (2010). The Economics Value of Volatility Timing using a Range-based Volatility Model. *Journal of Economic Dynamics & Control*, 34(11): 2288-2301.

[12] Chou, H.C., Zaabar, R., Wang, D. (2013). Measuring and Testing the Long-term Impact of Terrorist Attacks on the US Futures Market. *Applied Economics*, 45(2): 225-238.

[13] Cummins, J.D., Dionne, G., McDonald, J. B. (1990). Applications of the GB2 Family of Distributions in Modelling Insurance Loss Processes. *Insurance: Mathematics and Economics*, 9(4): 257-272.

[14] Godambe, V.P. (1985). The Foundations of Finite Sample Estimation in Stochastic Processes. *Biometrika*, 72(2): 419-428.

[15] Kumar, D. (2015). Sudden changes in extreme value volatility estimator: Modeling and forecasting with economics significance analysis. *Economic Modelling*, 49, 354-371.

[16] Li, D.X., Turtle, H. J. (2000). Semiparametric ARCH Models: An Estimating Function Approach. *Journal of Business & Economic Statistics*, 18(2): 174-186.

[17] Li, Hongquan, Hong, Yongmiao (2011). Financial Volatility Forecasting with Range-based Autoregressive Volatility Model. *Finance Research Letters*, 8(2): 69-76.

[18] Lin, E.M.H., Chen, C.W.S., Gerlach, R. (2012). Forecasting Volatility with Asymmetric Smooth Transition Dynamic Range Models. *International Journal of Forecasting*, 28(2): 384-399.

[19] Liang, Y., Thavaneswaran, A., Abraham, B. (2011). Joint Estimation using Quadratic Estimating Function. *Journal of Probability and Statistics*, doi.org/10.1155/2011/372512.

[20] Ng, Kok-Haur, Shelton, P., Thavaneswaran, A., Kooi-Huat, Ng (2015) Modelling the Risk or Price Durations in Financial Markets: Quadratic Estimating Functions and Applications. *Economic Computation and Economic Cybernetics Studies and Research*, 49(1): 223-238.

[21] Sin, C.Y. (2013). Using CARRX Models to Study Factors Affecting the Volatilities of Asian Equity Markets. *North American Journal of Economics and Finance*, 26: 552-564.

**Table 1.** Results for the CARR (1,1) model with standardized GB2 ($a = 1.0, p = 1.0, q = 2.0$) distribution and true parameters $\omega = 0.2$, $\alpha_1 = 0.3$, $\beta_1 = 0.4$ and $\psi_1 = 0.5$

| Sample Size, $T$ | | $\hat{\omega}$ | | | $\hat{\alpha}_1$ | | | $\hat{\beta}_1$ | | |
|---|---|---|---|---|---|---|---|---|---|---|
| | | CEF | LEF | ML | CEF | LEF | ML | CEF | LEF | ML |
| $T = 500$ | Mean | **0.2120** | 0.2142 | 0.2133 | **0.3054** | 0.3145 | 0.3061 | 0.3771 | 0.3647 | **0.3833** |
| | Bias | **0.0120** | 0.0142 | 0.0133 | **0.0054** | 0.0145 | 0.0061 | -0.0229 | -0.0353 | **-0.0167** |
| | SD | **0.0685** | 0.0737 | 0.0710 | 0.1012 | 0.1308 | **0.0928** | 0.1480 | 0.1604 | **0.1425** |
| | RMSE | **0.0693** | 0.0750 | 0.0723 | 0.1015 | 0.1315 | **0.0930** | 0.1497 | 0.1642 | **0.1435** |
| | SE | **0.0789** | 0.1015 | - | 0.1054 | 0.1338 | - | 0.1676 | 0.2129 | - |
| $T = 1000$ | Mean | 0.2067 | 0.2128 | **0.2061** | 0.3029 | 0.3063 | **0.2985** | 0.3872 | 0.3726 | 0.3957 |
| | Bias | 0.0067 | 0.0128 | **0.0061** | 0.0029 | 0.0063 | **-0.0015** | -0.0128 | -0.0274 | -0.0043 |
| | SD | 0.0514 | 0.0628 | **0.0458** | 0.0732 | 0.1009 | **0.0639** | 0.1130 | 0.1373 | 0.0939 |
| | RMSE | 0.0512 | 0.0640 | **0.0462** | 0.0735 | 0.1010 | **0.0639** | 0.1136 | 0.1400 | 0.0940 |
| | SE | 0.0549 | 0.0754 | - | 0.0767 | 0.1009 | - | 0.1182 | 0.1594 | - |
| $T = 1500$ | Mean | 0.2049 | 0.2096 | **0.2030** | 0.3023 | 0.3069 | **0.2996** | 0.3904 | 0.3780 | **0.3973** |
| | Bias | 0.0049 | 0.0096 | **0.0030** | 0.0023 | 0.0069 | **-0.0004** | -0.0096 | -0.0220 | **-0.0027** |
| | SD | 0.0428 | 0.0548 | **0.0362** | 0.0609 | 0.0855 | **0.0519** | 0.0946 | 0.1211 | **0.0739** |
| | RMSE | 0.0431 | 0.0556 | **0.0364** | 0.0609 | 0.0858 | **0.0519** | 0.0951 | 0.1231 | **0.0739** |
| | SE | 0.0447 | 0.0615 | - | 0.0631 | 0.0854 | - | 0.0969 | 0.1313 | - |
| $T = 2000$ | Mean | 0.2037 | 0.2061 | **0.2024** | 0.3014 | 0.3054 | **0.3014** | 0.3928 | 0.3855 | **0.3964** |
| | Bias | 0.0037 | 0.0061 | **0.0024** | 0.0014 | 0.0054 | **0.0014** | -0.0073 | 0.0145 | **-0.0036** |
| | SD | 0.0375 | 0.0490 | **0.0300** | 0.0549 | 0.0777 | **0.0441** | 0.0840 | 0.1099 | **0.0620** |
| | RMSE | 0.0376 | 0.0494 | **0.0301** | 0.0549 | 0.0779 | **0.0441** | 0.0843 | 0.1108 | **0.0622** |
| | SE | 0.0391 | 0.0586 | - | 0.0554 | 0.0762 | - | 0.0850 | 0.1171 | - |

**Table 2.** Results for the CARR (1,1) model with standardized Lognormal $(\sigma=0.5)$ distribution and true parameters $\omega=0.2$, $\alpha_1=0.3$, $\beta_1=0.4$ and $\psi_1=0.5$

| Sample Size, $T$ | | $\hat{\omega}$ | | | $\hat{\alpha}_1$ | | | $\hat{\beta}_1$ | | |
|---|---|---|---|---|---|---|---|---|---|---|
| | | CEF | LEF | ML | CEF | LEF | ML | CEF | LEF | ML |
| $T=500$ | Mean | **0.2097** | 0.2106 | 0.2113 | **0.2979** | 0.2974 | 0.2976 | **0.3873** | 0.3860 | 0.3850 |
| | Bias | **0.0097** | 0.0106 | 0.0113 | **-0.0021** | -0.0026 | -0.0024 | **-0.0127** | -0.0140 | -0.0150 |
| | SD | **0.0547** | 0.0582 | 0.0548 | **0.0477** | 0.0502 | 0.0485 | **0.1050** | 0.1118 | 0.1055 |
| | RMSE | **0.0555** | 0.0592 | 0.0560 | **0.0478** | 0.0502 | 0.0486 | **0.1057** | 0.1126 | 0.1066 |
| | SE | **0.0544** | 0.0571 | - | **0.0483** | 0.0505 | - | **0.1051** | 0.1102 | - |
| $T=1000$ | Mean | 0.2053 | 0.2075 | **0.2050** | 0.2993 | **0.2997** | 0.2993 | 0.3926 | 0.3887 | **0.3929** |
| | Bias | 0.0053 | 0.0075 | **0.0050** | -0.0007 | **-0.0003** | -0.0007 | -0.0074 | -0.0113 | **-0.0071** |
| | SD | **0.0364** | 0.0396 | 0.0366 | **0.0338** | 0.0353 | 0.0338 | 0.0714 | 0.0762 | **0.0714** |
| | RMSE | **0.0368** | 0.0403 | 0.0369 | **0.0338** | 0.0353 | 0.0338 | 0.0718 | 0.0770 | **0.0717** |
| | SE | **0.0367** | 0.0397 | - | **0.0342** | 0.0359 | - | 0.0733 | 0.0772 | - |
| $T=1500$ | Mean | 0.2038 | 0.2042 | **0.2036** | 0.2997 | **0.2998** | 0.2994 | 0.3944 | 0.3937 | **0.3947** |
| | Bias | 0.0038 | 0.0042 | **0.0036** | -0.0003 | **-0.0002** | -0.0006 | -0.0056 | -0.0063 | **-0.0053** |
| | SD | **0.0298** | 0.0310 | 0.0299 | **0.0274** | 0.0285 | 0.0275 | **0.0577** | 0.0602 | **0.0577** |
| | RMSE | **0.0300** | 0.0313 | 0.0301 | **0.0274** | 0.0285 | 0.0275 | **0.0580** | 0.0605 | **0.0580** |
| | SE | **0.0305** | 0.0320 | - | **0.0280** | 0.0293 | - | **0.0596** | 0.0624 | - |
| $T=2000$ | Mean | **0.2031** | 0.2033 | 0.2033 | **0.2999** | 0.2998 | 0.3000 | **0.3955** | 0.3950 | 0.3949 |
| | Bias | **0.0031** | 0.0033 | 0.0033 | **-0.0001** | -0.0002 | 0.0000 | **-0.0045** | -0.0050 | -0.0051 |
| | SD | **0.0254** | 0.0265 | 0.0264 | **0.0235** | 0.0247 | **0.0235** | **0.0494** | 0.0519 | 0.0506 |
| | RMSE | **0.0255** | 0.0267 | 0.0266 | **0.0235** | 0.0247 | **0.0235** | **0.0496** | 0.0521 | 0.0508 |
| | SE | **0.0263** | 0.0276 | - | **0.0242** | 0.0254 | - | **0.0514** | 0.0539 | - |

**Table 3.** Summary statistics for the AORD range data

| | |
|---|---|
| Mean | 0.9829 |
| Median | 0.8763 |
| Variance | 0.2558 |
| Skewness | 2.17 |
| Kurtosis | 13.86 |
| Minimum | 0.0409 |
| Maximum | 6.9083 |
| Ljung-Box, $Q_6$ | 968.568 |
| Ljung-Box, $Q_{12}$ | 1535.56 |

**Table 4.** Results for CARR (1,1) models using the ML, LEF and CEF methods. Values in parentheses are standard errors of parameter estimates

| Coefficient | **ML** | **LEF** | **CEF** |
|---|---|---|---|
| $\hat{\omega}$ | 0.0358 (0.0087) | 0.0386 (0.0102) | 0.0354 (0.0073) |
| $\hat{\alpha}_1$ | 0.1569 (0.0161) | 0.1589 (0.0178) | 0.1521 (0.0130) |
| $\hat{\beta}_1$ | 0.8065 (0.0207) | 0.8016 (0.0234) | 0.8117 (0.0170) |
| $\hat{a}$ | 1.4536 (0.4478) | - | - |
| $\hat{p}$ | 5.2471 (2.6898) | - | - |
| $\hat{q}$ | 6.7659 (4.9471) | - | - |
| $\hat{\sigma}_\varepsilon$ | 0.4289* | 0.4287* | 0.4286 |
| In-sample RMSPE | **0.4385** | **0.4385** | 0.4386 |
| In-sample MAPE | 0.3345 | **0.3245** | **0.3245** |
| Out-of-sample RMSFE | 0.4011 | 0.4012 | **0.4005** |
| Out-of-sample MAFE | 0.3362 | 0.3365 | **0.3352** |

*The value of $\hat{\sigma}_\varepsilon$ is calculated from $\hat{\varepsilon}_t = r_t / \hat{\psi}_t$

**Table 5.** Results for CARR models based on the CEF method. Values in parentheses are standard errors of parameter estimates

| Coefficient | **CARR (1,2)** | **CARR (2,1)** | **CARR (2,2)** |
|---|---|---|---|
| $\hat{\omega}$ | 0.0346 (0.0079) | 0.0317 (0.0075) | 0.0520 (0.0329) |
| $\hat{\alpha}_1$ | 0.1466 (0.0212) | 0.1669 (0.0216) | 0.1396 (0.0203) |
| $\hat{\alpha}_2$ |  | -0.0269 (0.0263) | 0.0803 (0.1307) |
| $\hat{\beta}_1$ | 0.8694 (0.1589) | 0.8276 (0.0205) | 0.4155 (0.8979) |
| $\hat{\beta}_2$ | -0.0515 (0.1378) |  | 0.3111 (0.7309) |
| $\hat{\sigma}_\varepsilon$ | 0.4267 | 0.4283 | 0.4264 |
| In-sample RMSPE | **0.4385** | 0.4386 | 0.4384 |
| In-sample MAPE | **0.3242** | 0.3246 | 0.3240 |
| Out-of-sample RMSFE | 0.4009 | **0.3998** | 0.4020 |
| Out-of-sample MAFE | 0.3359 | **0.3332** | 0.3373 |

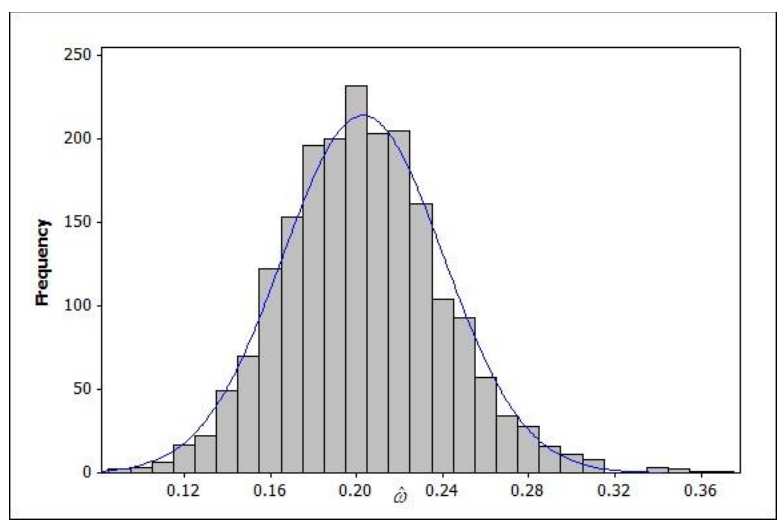

**Figure 1.** The histogram for the $\hat{\omega}$ obtained by CEF method ($\omega = 0.2, T = 2000$)

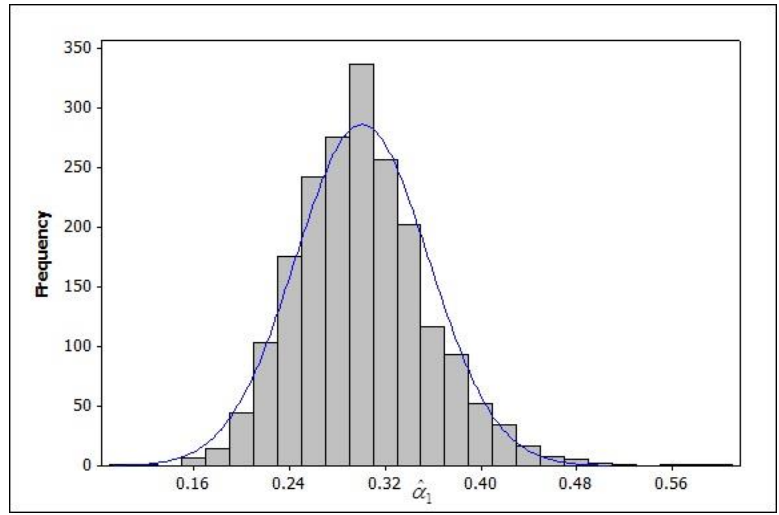

**Figure 2.** The histogram for the $\hat{\alpha}_1$ obtained by CEF method ($\alpha_1 = 0.3, T = 2000$)

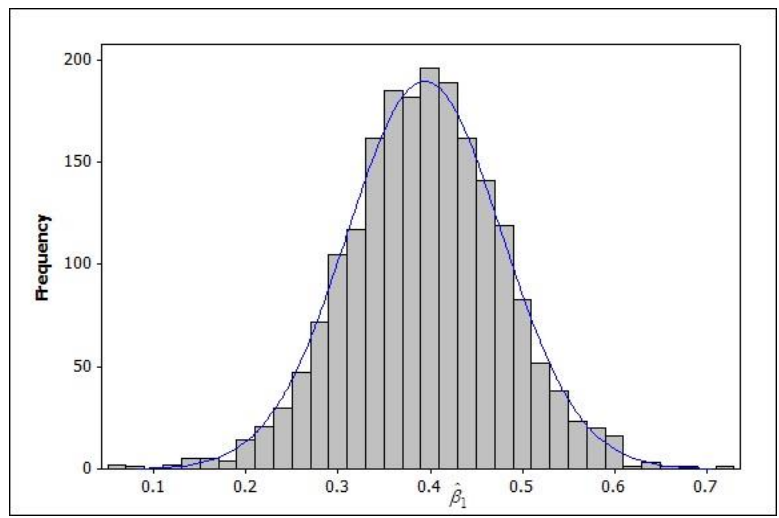

**Figure 3.** The histogram for the $\hat{\beta}_1$ obtained by CEF method ($\beta_1 = 0.4, T = 2000$)

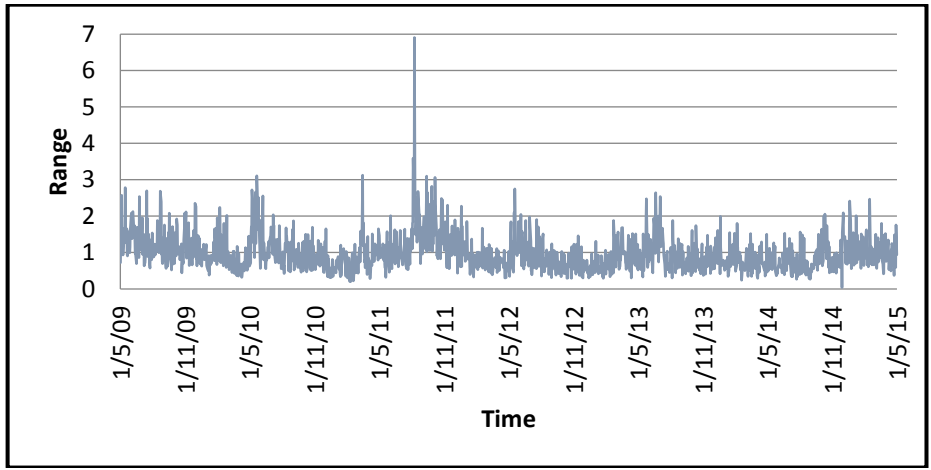

**Figure 4.** Time series plot of the AORD range series, $r_t$

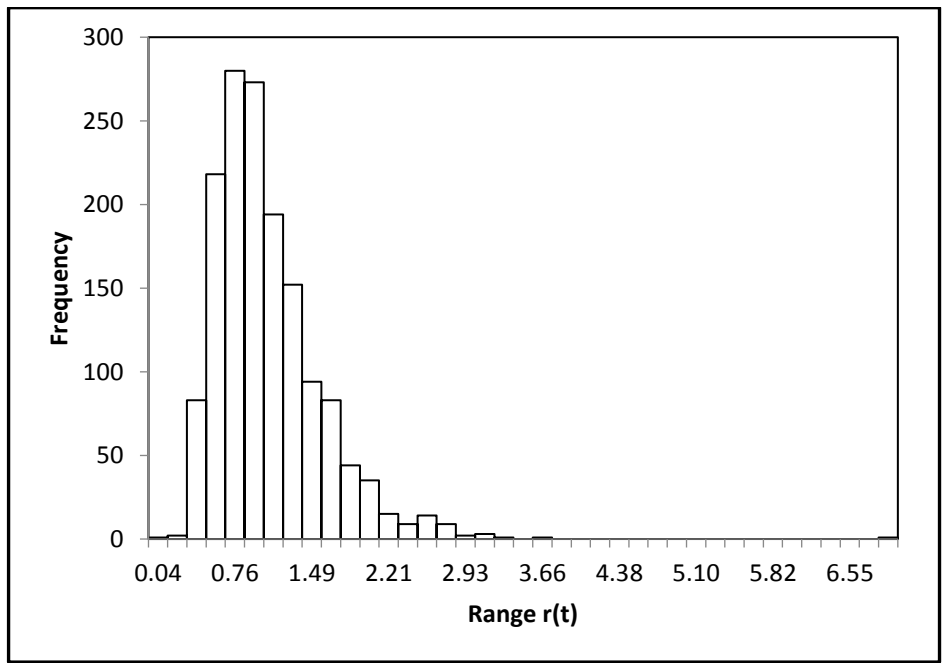

**Figure 5.** Histogram of the AORD range series, $r_t$

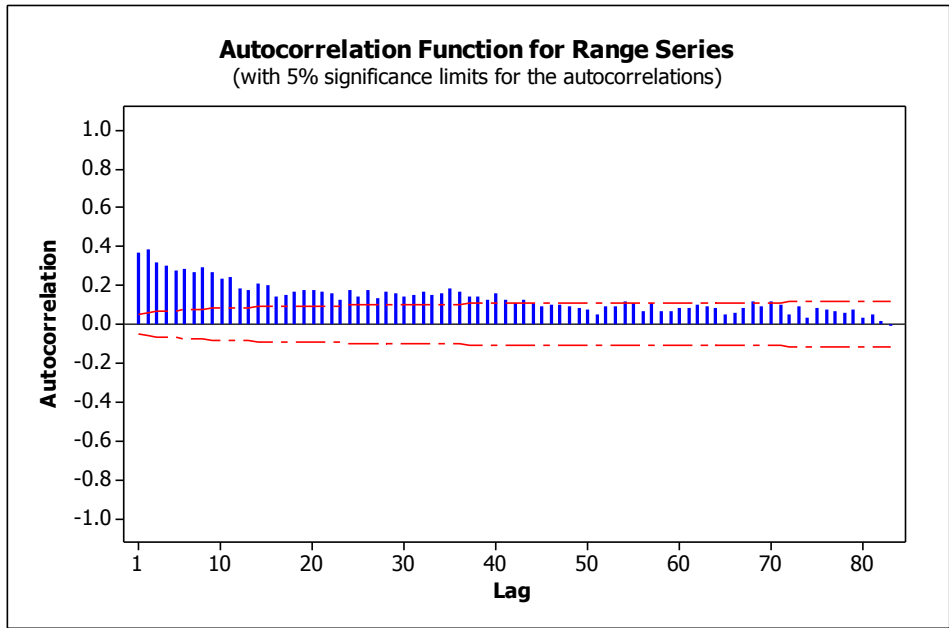

**Figure 6.** Autocorrelation function (ACF) of the AORD range series, $r_t$

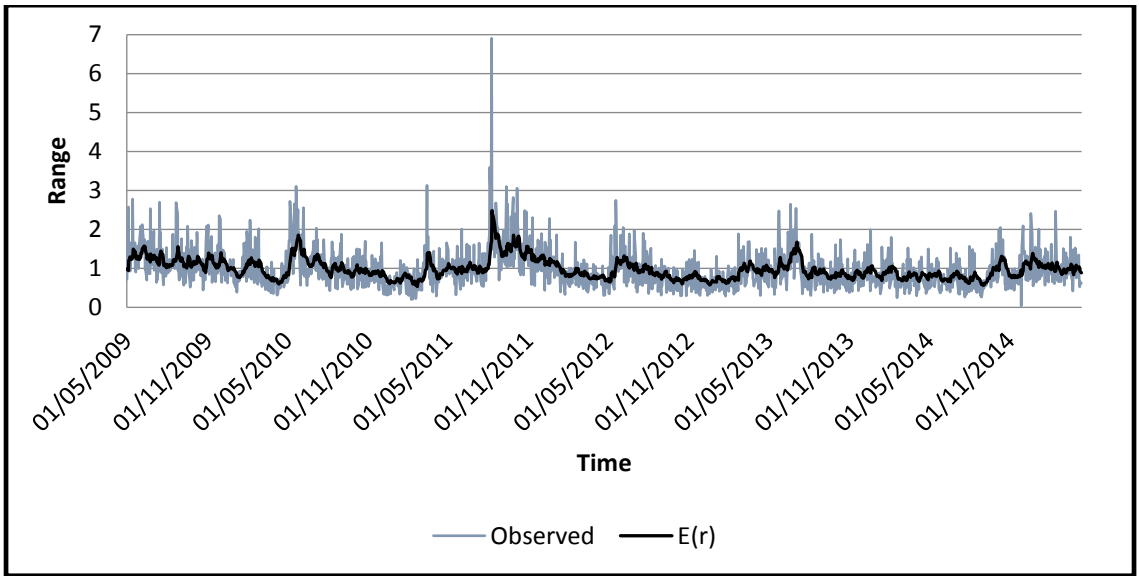

**Figure 7.** Observed $r_t$ and expected $E(r_t)$ using CARR (1,1) model and CEF method

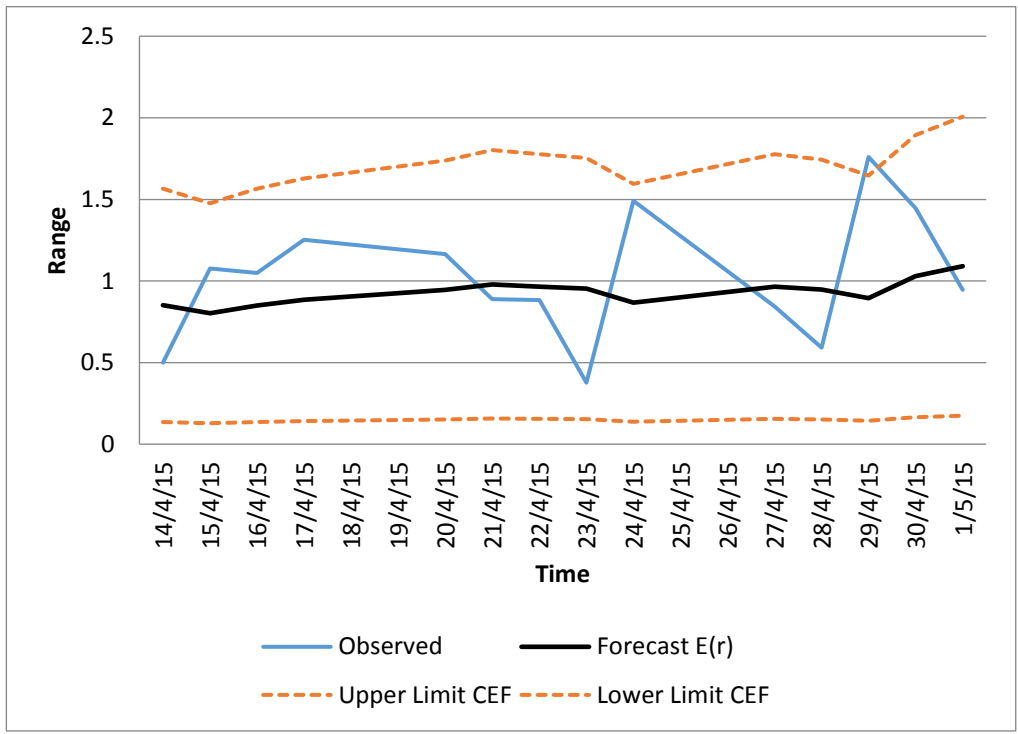

**Figure 8.** Observed $r_t$, forecast $\hat{r}_t$ and 95% confidence limits using CARR (1,1) model and CEF method